%% file: 0_main.tex
\documentclass[sigconf]{acmart}
\AtBeginDocument{%
  \providecommand\BibTeX{{%
    \normalfont B\kern-0.5em{\scshape i\kern-0.25em b}\kern-0.8em\TeX}}}

\usepackage{multirow}
\usepackage{multicol}
\usepackage{pdflscape}
\usepackage{makecell}
\usepackage{color}
\usepackage{enumitem}
\setlist[itemize]{leftmargin=*}

\newtheorem{thm}{Prompt}[section]

\copyrightyear{2024}
\acmYear{2024}
\setcopyright{acmlicensed}\acmConference[WWW '24 Companion]{Companion Proceedings of the ACM Web Conference 2024}{May 13--17, 2024}{Singapore, Singapore}
\acmBooktitle{Companion Proceedings of the ACM Web Conference 2024 (WWW '24 Companion), May 13--17, 2024, Singapore, Singapore}
\acmDOI{10.1145/3589335.3651232}
\acmISBN{979-8-4007-0172-6/24/05}

\begin{document}

\title{FashionReGen: LLM-Empowered Fashion Report Generation}
\author{Yujuan Ding}
\email{dingyujuan385@gmail.com}
\affiliation{
  \institution{The Hong Kong Polytechnic University, HK SAR}}

\author{Yunshan Ma}
\email{yunshan.ma@u.nus.edu}
\affiliation{
  \institution{National University of Singapore, Singapore}}

\author{Wenqi Fan}
\email{wenqifan03@gmail.com}
\authornote{Corresponding author: Wenqi Fan, Department of Computing, and 
 Department of Management and Marketing, The Hong Kong Polytechnic University.}
\affiliation{
  \institution{The Hong Kong Polytechnic University, HK SAR}}
  
\author{Yige Yao}
\email{yi-ge.yao@connect.polyu.hk}
\affiliation{
  \institution{The Hong Kong Polytechnic University, HK SAR}}

\author{Tat-Seng Chua}
\email{dcscts@nus.edu.sg}
\affiliation{
  \institution{National University of Singapore, Singapore}}
  
\author{Qing Li}
\email{csqli@comp.polyu.edu.hk}
\affiliation{
  \institution{The Hong Kong Polytechnic University, HK SAR}}

\begin{abstract}
Fashion analysis refers to the process of examining and evaluating trends, styles, and elements within the fashion industry to understand and interpret its current state, generating fashion reports. It is traditionally performed by fashion professionals based on their expertise and experience, which requires high labour cost and may also produce biased results for relying heavily on a small group of people. In this paper, to tackle the \textbf{Fashion} \textbf{Re}port \textbf{Ge}neration (\textbf{FashionReGen}) task, we propose an intelligent \textbf{F}ashion \textbf{A}nalyzing and \textbf{R}eporting system based the advanced Large Language Models (LLMs), debbed as \textbf{GPT-FAR}. Specifically, it tries to deliver FashionReGen based on effective catwalk analysis, which is equipped with several key procedures, namely, catwalk understanding, collective organization and analysis, and report generation. By posing and exploring such an open-ended, complex and domain-specific task of FashionReGen, it is able to test the general capability of LLMs in fashion domain. It also inspires the explorations of more high-level tasks with industrial significance in other domains. Video illustration and more materials of GPT-FAR can be found in \url{https://github.com/CompFashion/FashionReGen}. 
\end{abstract}


\vspace{-0.15in}

\maketitle

\input{1_introduction}
\input{2_method}

\input{3_results}
\input{4_conclusion}
\begin{acks}
The research described in this paper has been partly supported by the National Natural Science Foundation of China (project no. 62102335), General Research Funds from the Hong Kong Research Grants Council (project no. PolyU 15200021, 15207322, and 15200023), internal research funds from The Hong Kong Polytechnic University (project no. P0036200, P0042693, P0048625), Research Collaborative Project no. P0041282, and SHTM Interdisciplinary Large Grant (project no. P0043302), PolyU Distinguished Postdoctoral Fellowship (project no. P0048752). 
\end{acks}

\bibliographystyle{ACM-Reference-Format}
\bibliography{bib-refs}
\newpage

\end{document}

%% file: 1_introduction.tex
\section{Introduction}
The fashion industry is a vital component of global economy, characterized by its constant pursuit of novelty and changes. It is essential for fashion practitioners, enthusiasts and consumers to capture these shifts to get ahead of the right trends and make confident decisions~\cite{ma2020knowledge,ding2021leveraging,ding2023personalized}. 
To this end, specialized consulting and forecasting companies conduct seasonal or annual analysis, producing insightful fashion reports\footnote{www.wgsn.com; https://www.mckinsey.com/industries/retail/our-insights/state-of-fashion}. However, such \textbf{manual} production usually relies on a small group of experts to review, comprehend, summarize, create visual aids such as charts, analyze and write the reports. The manual process involves  high labor and intellectual cost. 

Developing an advanced approach to automate parts of or even the entire fashion report generation process is of great value. We term the task as \textbf{Fashion} \textbf{Re}port \textbf{Gen}eration (\textbf{FashionReGen}), which has been barely explored because of the challenges in different stages. First, it requires fine-grained recognition of categories, attribute details, and even fashion-specialized concepts for high-level comprehension of fashion content. Second, domain-specific knowledge and experiences are required to perform professional statistical analysis. Finally, proper selection and combination of multi-modal information are needed to integrate text, charts and images into a coherent, complete and insightful report. Even though previous deep learning techniques have achieved enormous progress in image recognition and text generation, they are mostly developed for the general domain while far from achieving expert-level performance in specific domains such as fashion. Fortunately, the breakthrough of Large Language Models (LLMs)~\cite{cheng2023gpt} shed light on this challenging problem, owing to the superb capability in multi-modal content comprehension and generation.

\begin{figure*}
    \centering
    \includegraphics[width=0.85\textwidth]{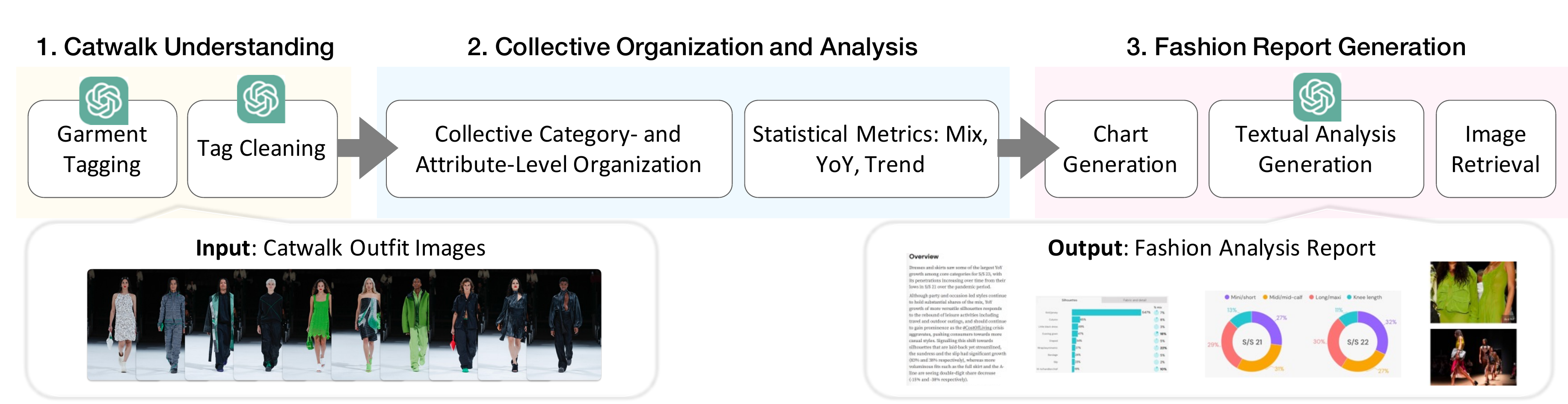}
    \caption{Diagram of our GPT-FAR system.}
\end{figure*}
To this end, we propose an intelligent \textbf{F}ashion \textbf{A}nalyzing and \textbf{R}eporting (\textbf{FAR}) system empowered by GPT models (named as \textbf{GPT-FAR}) for FashionReGen. It is an intelligent, also first-of-its-kind pipeline that enables automatic fashion analysis and report generation. Our current FashionReGen is based on catwalk analysis since catwalk is the place where high brands present their new designs. It tends to lead the trends and holds great impact on downstream fashion industry. The proposed GPT-FAR includes the processes of catwalk understanding, collective organization and analysis, and multi-modal fashion report generation. To facilitate the end-to-end FashionReGen based on catwalk observation, we first design a GPT-4V-based tagger to interpret the garments in the catwalk images, followed by a two-stage tag cleaner. Then, we employ professional metrics and charts to perform collective analysis. Finally, we develop an algorithm based on GPT-4V to generate textual analysis for the report with in-context learning based on statistical charts produced previously. 
Our contributions are:
\begin{itemize}
    \item We present FashionReGen, a high-level domain-specific task with significant research and application value. 
    \item We propose GPT-FAR, a system for automatic fashion report generation based on effective catwalk observation and analysis. 
    \item We develop a complete platform to enable users to generate their own fashion reports on specific collections.
\end{itemize}

%% file: 2_method.tex
\section{GPT-FAR System}
Our GPT-FAR consists of three major stages: 1) catwalk understanding, 2) collective analysis, and 3) fashion report generation.

\subsection{Catwalk Understanding}
\subsubsection{Garment tagging with GPT-4V}

We device an advanced \textbf{garment tagger} based on GPT-4V given its superior capability in image understanding~\cite{yang2023dawn}. 
For each catwalk image, the garment tagger first classifies each garment into one of the following categories: \textit{dresses}, \textit{skirts}, \textit{jackets}, \textit{coats}, \textit{trousers}, \textit{shorts}, \textit{knit and jersey}, \textit{sweater}, \textit{top}, \textit{blouses and woven tops}. The follow-up attribute-level tagging 
is strictly based on the methodology employed by WGSN\footnote{WGSN is professional fashion analysis company}. Specifically, the tagger is required to generate keywords to describe each garment for its \textit{style}, \textit{silhouette}, \textit{neckline}, \textit{length}, \textit{print and pattern}, \textit{detail}, \textit{embellishment} and \textit{fabric} if applicable. Such a tagging procedure allows to generate systematic and comprehensive descriptive tags for each garment, which is achieved with the following prompt:
\begin{thm}
    Can you tag the outfit in the image at garment level, which including two main steps. 1. recognize garments in the image and label them with categories from the category set of \#CATEGORY LIST and 2. for each garment, tag it from the the following aspects: \#ASPECT LIST. You have to list as MANY tags as possible suitable for each aspect, not only one. Report the tagging results for each garment with a single line pattern, following the corresponding category. Do not output anything other than the category and tags for the mentioned aspects. Here is an example for your tagging, <image: \{Category: Top; Style: Layered, Modern, ... Silhouette: Relaxed, ...; ...\}, \{Category: Skirt; Style: Casual, Street, ...; ... \}>.
\end{thm}

\subsubsection{GPT-produced tag cleaning}
Based on our experimental observation, the garment tagger is able to generate accurate, rich, and professional tags for target aspects. However, since the task is open-ended, it is difficult to strictly restrict the format of the output, even when we have emphasized the output format or shown examples in the prompt. A more tricky challenge is that it may generate different words meaning almost exactly the same attribute since the corpus is not and cannot be pre-defined. This is also a general issue of GPT-generated content, which is not friendly for subsequent processing, particularly statistical analysis. 

\begin{figure}[!t]
    \centering
    \includegraphics[width = 0.85\linewidth]{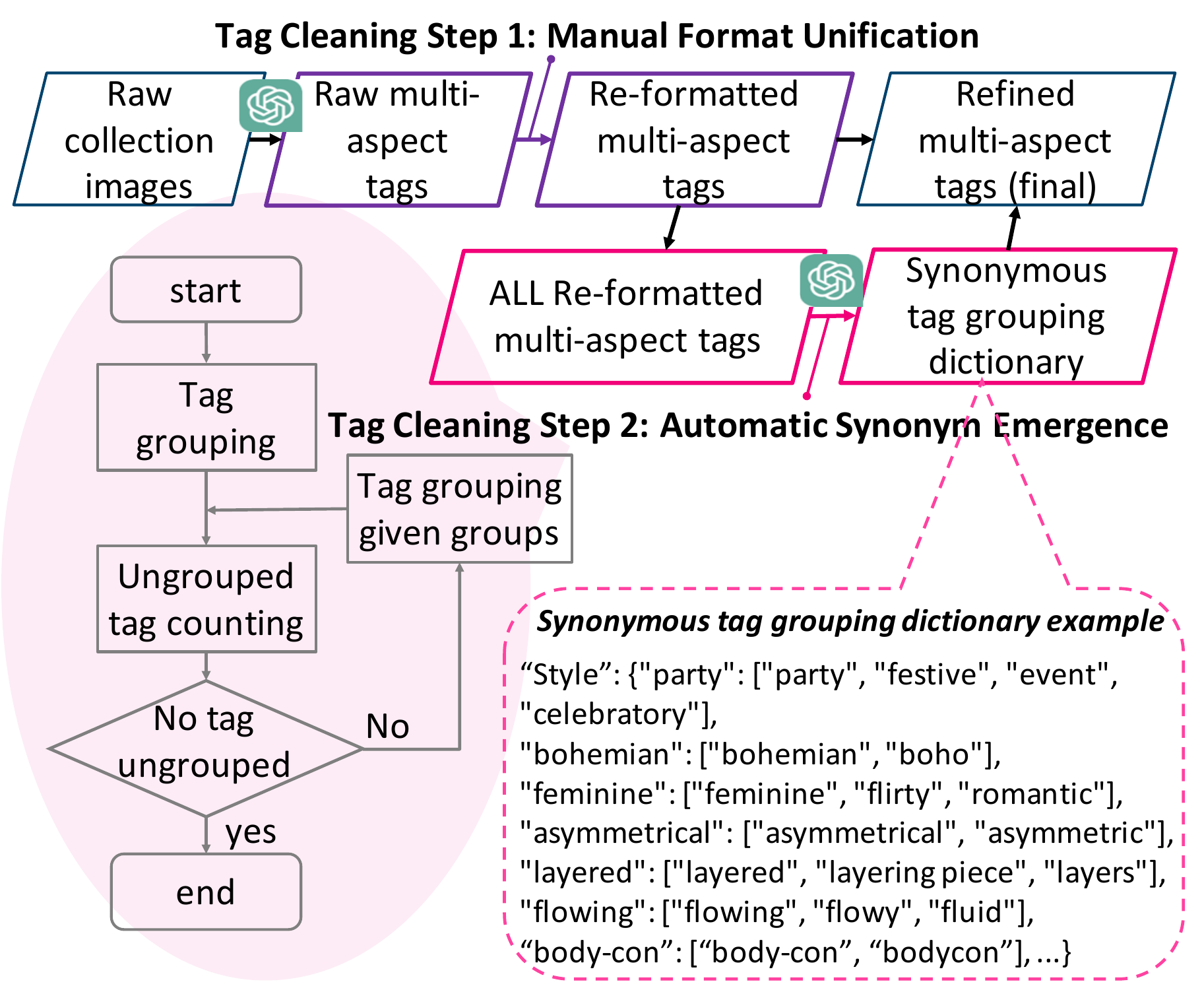}
    \caption{Data flow of the catwalk understanding process.}
    \label{fig:tagging_data}
\end{figure}

To address this issue, we design a two-stage tag cleaning strategy for our GPT-produced tags, including \textbf{manual format unification} and \textbf{automatic synonym emergence} (shown in Figure~\ref{fig:tagging_data}). For format unification, we set some rules to address the inconsistency issue in terms of capitalization, plurality, and other formats such as extra space. Synonym emergence is designed as an iterative grouping process on the whole tag corpus of specific aspect based on GPT-4. As shown in the bottom left in Figure~\ref{fig:tagging_data}, we apply a GPT agent to group 
synonym tags together with the following prompt: 
\begin{thm}
    You are an assistant to help group tagging corpus, specifically for the corpus describing \#attribute for \#CATEGORY. You are required to group similar words together and output a dictionary that map each word to the group? The corpus contains words as follows \#TAGS CORPUS. Make sure only words with almost the same meaning be grouped, NOT those describing the same aspect at a larger scale. Here is an example: 'draped': ['draped', 'draping', 'draped front', 'draped neckline', 'draped panel', 'draped shoulders', 'draped overlay', 'draped look’]. Only output the word groups as a dictionary, DO NOT output other descriptive or explanatory text!!
\end{thm}
We would expect that the guided agent would produce a perfect tag grouping dictionary. However, testing results suggest that so far it is challenging to achieve a one-step completion, especially when the tag corpus is relevantly large (with hundreds of identical tags). Therefore, we further design an iterative grouping process, which repeats the grouping operation for several times to ensure all tags with synonyms get well grouped. At each iteration, the GPT agent is required to group non-grouped tags into existing groups or create new groups with the following prompt. 
\begin{thm}
    ... Candidate word groups include \#TAG GROUPS. Check whether each word can be grouped into an existing group. If not, you can create new groups ...
\end{thm}
Note that this process groups only synonyms or different formats of words. We appreciate that GPT possesses expansive vocabulary to describe garment in detailed and professional words. 

\subsection{Collective Organization and Analysis}
Our fashion analysis is at collection-level, which requires
the organization of garment-level tagging results to obtain collective data. 
In specific, each garment $i$ is tagged by its category $c^i \in \mathcal{C}$ and the attribute details attached to the category $\{a^i\}_{a\in\mathcal{A}_a}$. The category set is $\mathcal{C} = \{c_1, c_2, ..., c_{\mathcal{C}|}\}$. The attribute set for the specific category $c$ is $\mathcal{A}^c = \{A^c_1, A^c_2, ..., A^c_{|\mathcal{A}^c|}\}$. Each attribute contains different values, for example, $A^c_1 = \{a^c_{11}, a^c_{12}, ...\}$, where $a^c_{11}$ denotes the specific attribute value of $A^c_1$. If the category \textit{dresses}, one corresponding attribute could be \textit{silhouette}, then under this attribute, the attribute value could be \textit{fit-and-flare}, \textit{layered}, \textit{etc}. 

Given a collection defined by its season (\textit{t}) and brand (\textit{b}) $\mathcal{S}_{t, b} = \{i_1, i_2, ..., i_{|\mathcal{S}_{t, b}|}\}$ (we may omit the subscripts for simplicity), the category-level index can be summarized as $\mathcal{D}^\mathcal{C}_\mathcal{S} = \{d^{c_1}_\mathcal{S}, d^{c_2}_\mathcal{S}, ..., d^{c_\mathcal|C|}_\mathcal{S}\}$, where $d^{c_1}_\mathcal{S}$ denotes the index of $\mathcal{S}$ of category $c_1$ and so on for others. Similarly, attribute-level index of collection $\mathcal{S}$ can be obtained. Taking the category $c$ as an example, the attribute index set is $\mathcal{A}^c$ and its index set is $\mathcal{D}^{\mathcal{A}^c}_\mathcal{S}=\{\mathcal{D}^{A^c_1}_{\mathcal{S}}, \mathcal{D}^{A^c_2}_{\mathcal{S}}, ... \}$. Each attribute-level index set can be summarized as $\mathcal{D}^{A^c} = \{d^{a^c_1}, d^{a^c_2}, ...\}$. 

We employ three key metrics for collective analysis, which are the mix ($Mix$), Year-on-Year index ($YoY$), and the list of evolving trend $T$. The category-level metrics (for category $c$) are as follows:
\begin{align}
    Mix^c_{\mathcal{S}_{t, b}} &= \frac{d^{c}_{\mathcal{S}_{t, b}}}{\sum_{\mathcal{D}^{c}_{\mathcal{S}_{t, b}}}} \times 100\%, \\
    YoY^c_{\mathcal{S}_{t, b}} &= \frac{Mix^c_{\mathcal{S}_{t, b}} - Mix^c_{\mathcal{S}_{t-1, b}}}{Mix^c_{\mathcal{S}_{t-1, b}}}. 
\end{align}

By replacing the category index $d_\mathcal{S}^{c, t}$ into attribute value index $d_{\mathcal{S}_{t, b}}^{a, t}$ and applying the corresponding attribute index set $\mathcal{D}_{\mathcal{S}_{t, b}}^A$, we can obtain the $Mix$ of attribute value $Mix^a_{\mathcal{S}_{t, b}}$ as well as the Year-on-Year index $YoY^a_{\mathcal{S}_{t, b}}$. In addition, combining yearly data of $Mix$ or $YoY$ can obtain corresponding evolving trend list $T$. 

\begin{table}[]
    \caption{Catwalk Dataset}
    \centering
    \scalebox{0.9}{
    \resizebox{0.46\textwidth}{!}{
        \begin{tabular}{m{0.8cm}m{3.5cm}m{3.4cm}}
         \toprule
           ~ &Current Version &Expandable Version \\
           \midrule
           Time Span & 2019\--2023 (5 years) & More years in the past or to come \\
           \midrule
           Brand &Chanel, Saint Laurent, LV, Valentino, Dior, Givenchy & More high fashion brands leading trends \\
           \midrule
           Seasons & Spring/Summer (SS) & Autume/Winter (AW), Pre-summer, Pre-fall\\
        \bottomrule   
        \end{tabular}} 
        }
    \label{tab:data}
\end{table}

\begin{figure*}
    \centering
    \includegraphics[width=0.9\textwidth]{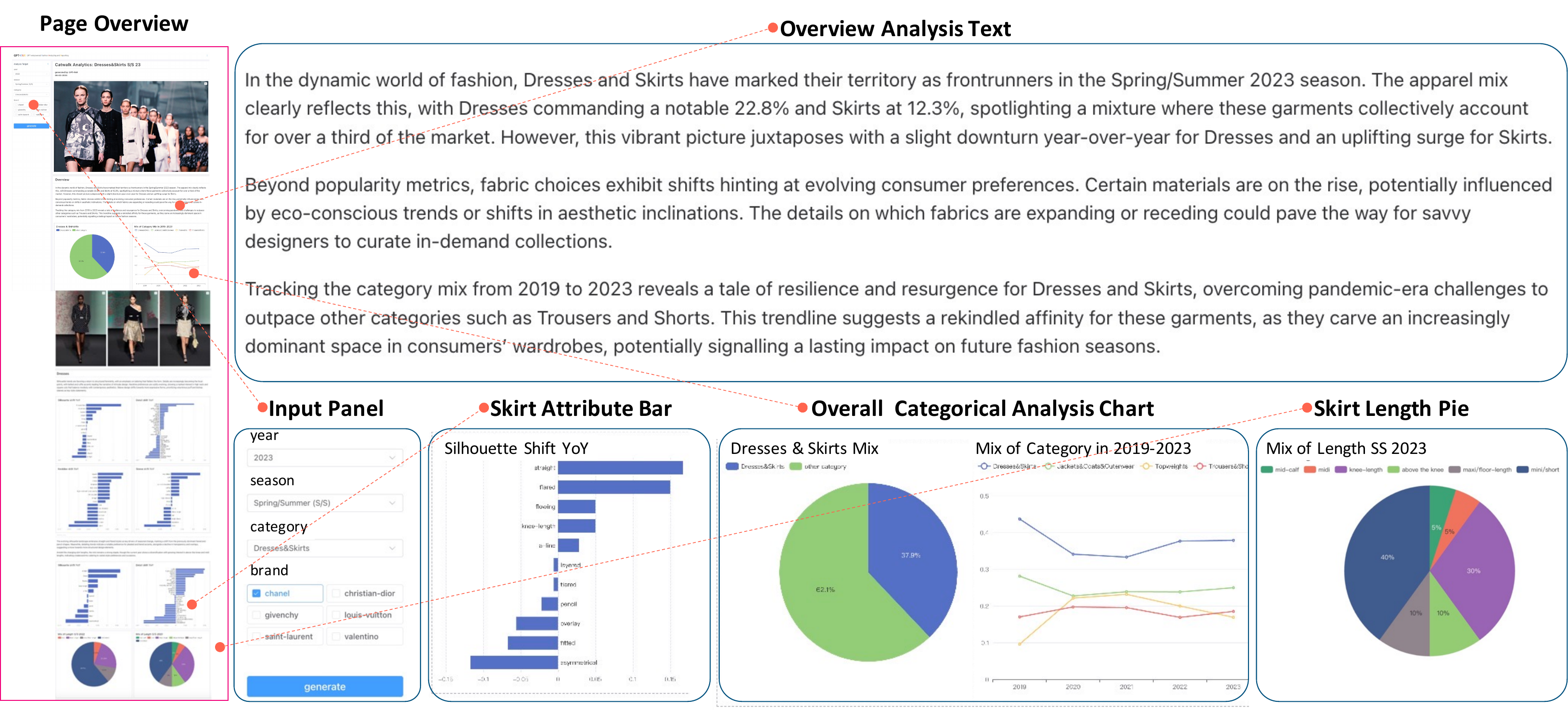}
    \caption{Illustration of the Fashion Analyzing and Reporting system and an example of the generated report.}
    \label{fig:page}
\end{figure*}

\subsection{Fashion Report Generation}
In order to make our fashion report instructive and insightful, the report to generate will be covering three main parts: charts for statistical analysis, catwalk images for illustration, and textural demonstration for state summary and highlights. In the current version, we simply plot charts based on the collective analytic data from previous stage and retrieve images relevant to the analyzed collection. For the textural analysis generation, we survey professional fashion reports and propose the chart-based generation method. 
The idea is to uncover the notable changeability or trends presented by charts, hopefully to produce meaningful insights. For overall textual analysis, we first design a GPT-4V-based agent with the following prompt:

\begin{thm}
    You are given several charts describing the fashion status specifically for \#REPORT TYPE of \#COLLECTION and SEASON. Each chart is about one specific aspect, e.g., fabric, silhouette. Try to generate several paragraphs (less than FIVE) in the format of an article. You are also given several examples of textual analysis based on charts as follows: \#EXAMPLEs. The length of the article should be around 250 characters. Do not use any key points or subtitles. 
\end{thm}

Furthermore, we present focused analysis in a short description for specific categories covered in the report, which is achieved by the following prompt:
\begin{thm}
    You are given several charts ... Try to generate a very short and neat piece of description (MUST less than two sentences) that can give an overview of the category or highlight the most significant trend. Please DO NOT make it too specific on specific aspects. You are also given several examples ... Please try to get the tone and style of the descriptions from the examples and apply then in your generation. 
\end{thm}

%% file: 3_results.tex
\section{Results and Analysis: A Case Study}
To evaluate the effectiveness of the proposed GPT-FAR system to deliver automatic FashionReGen, we collect catwalk images and review the generated reports based on them. Considering the computational and economic cost, the preliminary data is small amount  as shown in Table.~\ref{tab:data}.  
The system page of GPT-FAR, as shown in Figure~\ref{fig:page}, provides an input panel to select the analysis targets by year, season, category group and brand(s). Reports can be automatically generated by clicking the \textit{generate} button.

\subsection{Report Type and Structure}
Our report performs catwalk analytics on specific group of categories, such as \textit{Dresses \& Skirts}, \textit{Topweights}, \textit{Trousers \& Shorts}, \textit{Jackets \& Coats}. As a generated case shown in Figure~\ref{fig:page}, it is a multi-page report with a pre-defined layout. It starts with a cover page with the basic information, i.e., title, author and generating time of the report. The overall analysis page is next, which has a three-column layout with textual analysis in the left, charts lying vertically in the middle, and image illustrations in the right. The rest of the report is category-specific analysis. 

\subsection{Generation Results and Discussion}
We have the following observation and discussion on the report generation and the middle-stage results of GPT-FAR: 
1) GPT-4V enables to generate high-quality descriptive tags for garments owing to its visual understanding capacity. However, it also remains challenging to output only concise and clean tags, which makes the tag cleaning strategy critical for the further analytics on them;
2) The generated textual demonstration, especially for the overall analysis, is reasonable, fluent and professional; and 
3) Overall, GPT-FAR generates high-quality fashion reports, which are comprehensive, illustrative, and in a hybrid modality of presentation. It offers a platform for automatic fashion analysis and report generation. 

%% file: 4_conclusion.tex
\section{Conclusion}
The paper posed a meaningful domain-specific task: fashion report generation (FashionReGen) relying on data observation and analysis. Empowered by LLM advances, a novel Fashion Analyzing and Reporting framework (GPT-FAR) was proposed, which enables automatic FashionReGen with the joint efforts of GPT-series models and domain-specific knowledge. A report generation system was further developed based on GPT-FAR, 
offering the platform for public users to generate their own fashion reports. 

This work is still a preliminary demo, which has huge potential for further enhancement. In future, we will include more sources of data, such as social media, forum and business data, in order to offer more perspectives and opinions in the analysis. Meanwhile, the type of fashion reports can be expanded. We will also work on specific technical challenges to offer higher degree of automation and intelligence
. The evaluation of FashionReGen is another important direction to study in the future. 